\documentclass[3p, times, 12pt]{elsarticle}

\usepackage{footnote}
\makesavenoteenv{tabular}
\makesavenoteenv{table}
\usepackage{booktabs} 
\usepackage{color}
\usepackage[usenames,dvipsnames,svgnames,table]{xcolor}
\usepackage{listings}
\usepackage{hyperref}
\usepackage{amsmath}
\usepackage{amssymb}

\definecolor{urlcolor}{RGB}{15,25,222}
\hypersetup{
    bookmarks=true,         
    unicode=false,          
    pdftoolbar=true,        
    pdfmenubar=true,        
    pdffitwindow=false,     
    pdfstartview={FitH},    
    pdftitle={},    
    pdfauthor= {},     
    pdfsubject={N}, 
    pdfcreator={Poslavsky},
    pdfproducer={hz}, 
    pdfkeywords={}, %
    pdfnewwindow=true,      
    colorlinks=true,       
    linkcolor=black,          
    citecolor=black,        
    filecolor=magenta,      
    urlcolor=urlcolor           
}

\lstdefinelanguage{scala}{
  morekeywords={%
          abstract,case,catch,class,def,do,else,extends,%
          false,final,finally,for,forSome,if,implicit,import,lazy,%
          match,new,null,object,override,package,private,protected,%
          return,sealed,super,this,throw,trait,true,try,type,%
          val,var,while,with,yield},
  otherkeywords={=>,<-,<\%,<:,>:,\#,@},
  sensitive=true,
  morecomment=[l]{//},
  morecomment=[n]{/*}{*/},
  morestring=[b]",
  morestring=[b]',
  morestring=[b]"""
}[keywords,comments,strings]

\definecolor{commcolor}{RGB}{105,105,105}
\newcommand\Rings{\textsc{Rings{}} }
\newcommand\Singular{\textsc{Singular{}} }
\newcommand\Mathematica{\textsc{Mathematica{}} }

\newcommand\ZZ{\mathbb{Z}}
\newcommand\QQ{\mathbb{Q}}
\newcommand\GF{\mathbb{GF}}
\newcommand\Groebner{Gr{\"o}ebner{} }

\journal{Computer Physics Communications}
\begin{document}
\begin{frontmatter}
\title{\textsc{Rings}: an efficient Java/Scala library for polynomial rings}

\author[a]{Stanislav Poslavsky\corref{SP}}

\cortext[SP] {\textit{E-mail address:} 
\href{mailto:stvlpos@mail.ru}{stvlpos@mail.ru}}
\address[a]{Institute for high-energy physics NRC "Kurchatov Institute", 142281 Protvino, Russia}

\begin{abstract}
In this paper we briefly discuss \Rings --- an efficient lightweight library for commutative algebra. Polynomial arithmetic, GCDs, polynomial factorization and Gr{\"o}bner bases are implemented with the use of modern asymptotically fast algorithms. \Rings can be easily interacted or embedded in applications in high-energy physics and other research areas via a simple API with fully typed hierarchy of algebraic structures and algorithms for commutative algebra. The use of the Scala language brings a quite novel powerful, strongly typed functional programming model allowing to write short, expressive, and fast code for applications. At the same time Rings shows one of the best performances among existing software for algebraic calculations.
\end{abstract}

\begin{keyword}
computer algebra; commutative algebra; polynomial arithmetic; 
\end{keyword}
\end{frontmatter}

\noindent {\bf PROGRAM SUMMARY}

\begin{small}
\noindent
{\em Program Title:} \textsc{Rings}\\
{\em Licensing provisions:} Apache 2.0\\
{\em Programming language:} Java, Scala\\
{\em Computer:} Any\\
{\em Operating system:} Linux/MacOS/Windows\\
{\em No. of lines in distributed program, including test data, etc.:} 101,433\\
{\em Running time and RAM:} Problem dependent\\
{\em Nature of problem:} Fast methods for rational function arithmetic, simplification of polynomial expressions, \Groebner bases and other related computer algebra methods naturally arising in physical applications \\
{\em Solution method:} Efficient implementation of modern asymptotically fast algorithms in Java language \\
{\em External routines:} Java 8 and higher, Scala 2.11 or 2.12 \\
{\em Additional comments:} project page: \href{https://github.com/PoslavskySV/rings}{https://github.com/PoslavskySV/rings}, \\ 
\phantom{{\em Additional comments:}} documentation:  \href{http://rings.readthedocs.io/en/latest/}{http://rings.readthedocs.io/en/latest/}\\
\end{small}

\section{Introduction}
Efficient implementation of polynomial rings and related concepts is crucial for modern computational algebra. Computational research areas such as commutative algebra, algebraic geometry, and number theory are tightly coupled with practical calculations in modern physics. For example, rational-function arithmetic is a key component of nearly all symbolic computations in high-energy physics. More specialized algebraic concepts arise in, for example, computation of (multi)loop Feynman diagrams: recent developments in this area are heavily based on such tools as Gr{\"o}bner bases, multivariate residues \cite{Larsen:2015ped, Larsen:2017aqb}, Nullstellensatz certificates \cite{Meyer:2016slj}, modular methods, rational reconstruction \cite{vonManteuffel:2014ixa}, etc. It is also worth noting that in addition to physics, efficient methods for polynomial rings are becoming to play an important role for other quite applied areas like cryptography (see e.g.\ recent developments in homomorphic encryption \cite{FHE}) or even machine learning \cite{MLCA}.

Among the modern open source software there are only few tools which are able to perform routines such as multivariate polynomial GCD or factorization at a speed sufficient for challenging real-world problems. Existing tools with such functionality are commonly implemented as a computer-algebra systems and each has its own interactive interface and domain-specific programming language as the only means to interact with it (for example \Singular \cite{Singular} is one of the most comprehensive and performant open-source systems for such type of computations). Applications which internally use this kind of mathematics are dependent on an external large-scale system, and all the users are forced to have the installation of the system on their machines. This complicates portability and may be inconvenient. The above considerations serve as one of the motivations for the development of the new open-source high-performance lightweight \textit{library} (not a full-featured system) for polynomial arithmetic with a clean API, which meets modern standards of software development. Another, no less important motivation, was to achieve the best possible performance by using the the most recent asymptotically fast algorithms and highly optimized implementation.

Several specialized open-source libraries which meet the above concern exist. The well-known \textsc{NTL} \cite{ntl} and \textsc{FLINT} \cite{flint} C/C++ libraries show unmatched performance in various aspects of computational number theory, and are widely used in related applications (but they support only univariate polynomial arithmetic). In high-energy physics, the \textsc{GiNaC} \cite{GiNaC} C++ symbolic library and especially \textsc{FORM} \cite{FORM,Ruijl:2017dtg} (though a computer-algebra system, it is quite lightweight and easily interacted) are widely used for practical computations, and both have support of polynomial rings to some extent. However, the performance of polynomial methods such as GCD is still not satisfactory for applications heavily using polynomial arithmetic, so e.g.\ tools for multiloop computations (see \textsc{FIRE} \cite{Smirnov:2008iw} or \textsc{Reduze} \cite{vonManteuffel:2012np}) have to use more specialized closed-source mathematical libraries (in particular \textsc{Fermat} \cite{Fermat}, which is highly performant but misses implementation of some important mathematical concepts like \Groebner bases or polynomial factorization). \Rings covers all computational topics mentioned above along with many other mathematical concepts while giving quite high or even unmatched performance in many standard applications which arise in physics on routine basis.

\Rings is a cross-platform library written entirely in Java and thus fully compatible with any modern JVM-based language (Scala/Closure/Kotlin/Groovy etc.), or easily interacted with from native applications both on POSIX systems and Windows (either via Java native interface (C/C+) or via simple pipes). The Scala extension allows for expressive type-safe interaction with the library from within a custom Scala application, as well as from the REPL console. Both Java and Scala are strongly and statically typed languages, and the mathematical structures used in \Rings also form a fully typed hierarchy, in contrast to many computer-algebra systems and libraries that operate either with untyped or weakly typed objects or use duck typing\footnote{one example of a computer-algebra package with a comprehensive type system is \textsc{Axiom} \cite{Axiom}, which comes with its own statically typed domain-specific programming language developing from mid 60s}. This strong typing is manifested e.g.\ in the fact that polynomials from different rings have different compile-time component types (example in Scala):
\begin{lstlisting}[language=scala]
// poly in %$\color{commcolor}\ZZ[x]$%
val polyZ : UnivariatePolynomial[IntZ] = ...
// poly in %$\color{commcolor}\QQ[x] \,\,\, (\QQ = Frac[\ZZ]$%)
val polyQ : UnivariatePolynomial[Rational[IntZ]] = ...
\end{lstlisting}
(in actual Scala code one can omit explicit type annotation in most places having them automatically inferred at compile time).

In the next section we give an overview of the most important \Rings features. Section~\ref{sec_bench} presents the benchmarks, Section~\ref{sec_impl} is dedicated to the algorithms and implementation aspects. All examples in this paper are given in Scala. Examples in Java can be found at the \Rings website \href{http://ringsalgebra.io}{http://ringsalgebra.io} and at \href{http://rings.readthedocs.io}{http://rings.readthedocs.io}. The source code of the library is hosted at GitHub \href{https://github.com/PoslavskySV/rings}{https://github.com/PoslavskySV/rings}.

\section{Overview}
In a single sentence, \Rings allows to construct different rings and perform arithmetic in them, including both very basic math operations and advanced methods like polynomial factorization, linear-systems solving and \Groebner bases. Below we'll illustrate these features in step-by-step examples.

Our starting point is to take some ground ring, for example a finite field $\GF(17,3)$, and perform some basic math in it:
\begin{lstlisting}[language=scala, numbers=left]
// Galois field %$\color{commcolor}\GF(17, 3)$%
implicit val gf = GF(17, 3, "t")
// parse ring element from its string representation
val t  = gf("t")
// or create element programmatically
val t1 = 1 + t.pow(2)
// do some basic math (+-*/) with elements of gf
val t2 = 3 + t1 - t.pow(22) / (1 + t + t1.pow(999))
\end{lstlisting}
It is worth emphasizing that while all explicit type annotations are usually omitted in the code, the Scala compiler will in fact infer them automatically, so the above lines are actually treated as:
\begin{lstlisting}[language=scala]
implicit val gf : Ring[UnivariatePolynomialZp64] = ...
val t           : UnivariatePolynomialZp64       = ...
val t1          : UnivariatePolynomialZp64       = ...
val t2          : UnivariatePolynomialZp64       = ...
\end{lstlisting}
Here the \texttt{Ring[E]} interface is a super type for all rings, so \texttt{Ring[UnivariatePolynomialZp64]} is a ring of elements of type \texttt{UnivariatePolynomialZp64}. The latter is used for univariate polynomials over $\ZZ_p$ --- that is the actual representation of elements of Galois fields (a more elaborate discussion of rings and polynomials is given in Section~\ref{sec_impl}).

In line 4, an element is parsed from its string representation. This line is actually a syntactic sugar for \texttt{gf.parse("t")}. In fact, there are several such methods for different conversions, which are all may be called in the same way (this is a common pattern in Scala):
\begin{lstlisting}[language=scala]
// from string
val elem = gf("1 + t^2")
// from Int
val unit = gf(1)
// from elements of other GF fields
val elementFromOtherField = GF(19, 5).randomElement()
val cast = gf(elementFromOtherField)
// synthetic sugar for multiple assignment
// val el1 = gf("t + 1"); val el2 = gf("t + 2")
val (el1, el2) = gf("t + 1", "t + 2")
\end{lstlisting}

Note that the ring \texttt{gf} in line 2 is defined \texttt{implicit}. This means that the Scala compiler will delegate all math operations on objects of type \texttt{UnivariatePolynomialZp64} to that implicit ring instance (with the use of Scala implicit conversions):
\begin{lstlisting}[language=scala]
t1 - t2 // is equivalent to gf.subtract(t1, t2)
t1 * t2 // is equivalent to gf.multiply(t1, t2)
3  + t1 // is equivalent to gf.add(gf.valueOf(3), t1)
\end{lstlisting}
With no \texttt{implicit} keyword specified, math operations will be performed as plain polynomial operations. For example, \texttt{(t1 + t2)} will be treated as \texttt{t1.add(t2)} instead of \texttt{gf.add(t1, t2)}: obviously, addition of elements in the Galois field is not the same as simple addition of polynomials.

The ground ring $\GF(17,3)$ was chosen only for our illustration purposes. In fact one can use any of the built-in rings absolutely in the same manner. The built-in rings include $\ZZ$, $\QQ$, $\QQ(i)$, $\ZZ_p$, $\GF(p,k)$ (with arbitrary large $p$), field extensions $F(\alpha_1, \dots, \alpha_s)$ (including algebraic number fields), $Frac(R)$, $R[x]$ and $R[\vec X]$ where $R$ is an arbitrary ground ring (which may be either one or any combination of the listed rings). \Rings uses a specifically optimized implementation for some particular rings in order to achieve the best performance. For example, the ring $\ZZ_p$ with $p$ less than $2^{64}$ (machine integer) has a different implementation than the same ring with larger $p$. Likewise, polynomials over $\ZZ_p$ with $p$ less than and greater than $2^{64}$ (like \texttt{UnivariatePolynomialZp64} from our example) also have different implementations.

Our next step it to define some multivariate polynomial ring over the ground ring $\GF(17,3)$. Below we define such ring and perform some math operations in the same fashion as we did above:
\begin{lstlisting}[language=scala, numbers=left, firstnumber=7]
// multivariate ring %$\color{commcolor}\GF(17, 3)[x, y, z]$%
implicit val ring = MultivariateRing(gf, Array("x", "y", "z"), GREVLEX)
val (x, y, z) = ring("x", "y", "z")
// construct some multivariate polynomials
val p1 = (t.pow(2) + 1)*x*y.pow(2)*z + (t + 1)*x.pow(5) * z*y.pow(6) + 1
val p2 = p1.pow(2) + (t + 1)*x.pow(2)*y.pow(2) + (t.pow(9) + 1)*z.pow(7)
\end{lstlisting}
Again, the \texttt{ring} instance is defined \texttt{implicit} so that all math operations with multivariate polynomials (which have type \texttt{MultivariatePolynomial[UnivariatePolynomialZp64]} in our example) will be delegated to that instance.

In line 8 we explicitly specified to use the \texttt{GREVLEX} monomial order for multivariate polynomials. This choice affects some algorithms like multivariate division and Gr{\"o}bner bases. The explicit order may be omitted (\texttt{GREVLEX} will be used by default). Several mathematical operations are especially important in the rings of polynomials:
\begin{lstlisting}[language=scala, numbers=left, firstnumber=14]
// standard division with remainder
val (div, rem) = p2 /%\%% p1
// define a set of polynomial generators
val (i1, i2, i3) = (x + y + z, x - y - z, y.pow(2) - z.pow(2))
// do multivariate division with remainder (polynomial reduction)
val (div1, div2, div3, rem1) = p2 /%\%%/%\%% (i1, i2, i3)
assert ( p2 == div1 * i1 + div2 * i2 + div3 * i3 + rem1 )
// construct Ideal from a set of generators 
// (Groebner basis with GREVLEX order will be automatically computed)
val ideal = Ideal(Seq(i1, i2, i3), GREVLEX)
// reduce poly modulo ideal
val p3 = p2 %\%%%\%% ideal
\end{lstlisting}
(a full list of operators defined in Scala (like those \texttt{/\%}, \texttt{/\%/\%}, \texttt{\%\%} used above) can be found in the  \href{http://rings.readthedocs.io/en/latest/guide.html#scala-dsl}{online manual}).

Simple division with remainder is supported for all Euclidean rings. This and other properties of rings may be checked by means of the following methods:
\begin{lstlisting}[language=scala]
ring.isEuclideanRing // whether ring is Euclidean ring
ring.isField         // whether ring is a field
ring.isFinite        // whether ring is finite
ring.cardinality     // ring cardinality (BigInteger)
ring.characteristic  // ring characteristic (BigInteger)
\end{lstlisting}

Polynomial greatest common divisors and polynomial factorization work for polynomials over all available built-in rings. Continue our example:
\begin{lstlisting}[language=scala, numbers=left, firstnumber=24]
// GCD of polynomials from %$\color{commcolor}\GF(17, 3)[x, y, z]$%
val gcd1 = PolynomialGCD(p1 * p3, p2 * p3)
assert ( gcd1 %\%% p3 === 0 ) 
val gcd2 = PolynomialGCD(p1 * p3, p2 * p3 + 1)
assert ( gcd2.isConstant ) 

// large polynomial from %$\color{commcolor}\GF(17, 3)[x, y, z]$%
// with more than %$\color{commcolor}4 \times 10^3$% terms and total degree of 145
val hugePoly = p1 * p2.pow(2) * p3.pow(3)
// factorize it
val factors = Factor(hugePoly)
\end{lstlisting}
This is one of the key features of the \Rings library: it does polynomial GCD and factorization of really huge polynomials over different ground rings robustly and fast. The performance achieved in Rings is one of the best among the existing software (see Section~\ref{sec_bench}).

To illustrate how the performance of e.g. polynomial GCD is manifested in applications, suppose we need to solve a system of linear equations with symbolic coefficients --- a task which arises frequently in many mathematical (e.g.\ sparse Hensel lifting or computation of Gr{\"o}bner bases) and physical (e.g.\ reduction of loop integrals) applications. It is worth noting that linear systems arising in, for example, the Passarino--Veltman method for expressing tensor one-loop integrals through scalar integrals may be quite huge (e.g.\ for a QCD process with four incoming and four outgoing quarks, one may need to reduce $367 \times 367$ matrices with non-trivial \textit{symbolic} coefficients). To solve such tricky systems in a field of rational functions one may try luck with direct methods (plug the system as is with rational coefficients) or use e.g.\ modular techniques \cite{vonManteuffel:2014ixa}, which implies switching from the $\QQ$ ground ring to some finite field (i.e.\ replacing all rational numbers with their images modulo some prime number). \Rings allows to solve linear systems over arbitrary fields quite efficiently, so one can try both approaches. In the continuation of our example, we use $\GF(17, 3)[x, y, z]$ for the ring of coefficients, so the solution belongs to the field of rational functions over 3 variables with coefficients from $\GF(17, 3)$. This can be accomplished in \Rings easily:

\begin{lstlisting}[language=scala, numbers=left, firstnumber=35]
// field of rational functions %$\color{commcolor}Frac(\GF(17, 3)[x, y, z])$%
implicit val ratRing = Frac(ring)
// convert x, y, z and t to elements of %$\color{commcolor}Frac(\GF(17, 3)[x, y, z])$%
// that is rationals with unit denominators
val (rx, ry, rz, rt) = ratRing(x, y, z, ring(t))
// lhs matrix
val lhs =
  Array(
    Array(rt + rx + rz, ry * rz, rz - rx * ry),
    Array(rx - ry - rt, rx / ry, rz + rx / ry),
    Array(rx * ry / rt, rx + ry, rz / rx + ry)
  )
// rhs column
val rhs = Array(rx, ry, rz)
// solve the system with Gaussian elimination
val solution = LinearSolver.solve[ratRing.ElementType](ratRing, lhs, rhs)
\end{lstlisting}
The standard Gaussian elimination uses $O(n^3)$ field operations, which means $O(n^3)$ multivariate polynomial GCDs (since each fraction should be reduced). The speed-up of the polynomial GCD by a factor two reduces the time to solve the entire system by nearly an order of magnitude --- that's why rational function arithmetic often becomes a bottleneck in such computations.

Line 40 is a syntactic sugar which converts elements of \texttt{ring} (multivariate polynomials) to elements of \texttt{ratRing} (rational functions), that is: 
\begin{lstlisting}[language=scala]
val rx = Rational(x, 1) // x is numerator and 1 is denominator
val ry = Rational(y, 1) // y is numerator and 1 is denominator
...
\end{lstlisting}
The type that will be inferred for variables \texttt{rx, ry, rz} and \texttt{rt} is:
\begin{lstlisting}[language=scala]
Rational[MultivariatePolynomial[UnivariatePolynomialZp64]]
\end{lstlisting}

Another important feature of \Rings is that it is a library which allows to write short and expressive code on top of it. Consider the following short example, which implements a solver for Diophantine equations --- that is a straightforward generalization of the extended GCD on more than two arguments (algorithm from Sec.~4.5 in \cite{CAHB}):
\begin{lstlisting}[language=scala, numbers=left, firstnumber=51]
/** 
 * Solves equation %$\color{commcolor} \sum f_i s_i  = gcd(f_1, \dots, f_N)$% for given %\color{commcolor}$f_i$% and unknown %\color{commcolor}$s_i$%
 * @return a tuple (gcd, solution)
 */
def solveDiophantine[E](fi: Seq[E])(implicit ring: Ring[E]) =
  fi.foldLeft((ring(0), Seq.empty[E])) { case ((gcd, seq), f) =>
    val xgcd = ring.extendedGCD(gcd, f)
    (xgcd(0), seq.map(_ * xgcd(1)) :+ xgcd(2))
  }
\end{lstlisting}
With this function it is very easy to implement, for example, an efficient algorithm for partial fraction decomposition with just a few lines of code. The resulting function will work with elements of arbitrary fields of fractions:
\begin{lstlisting}[language=scala, numbers=left, firstnumber=60]
/** Computes partial fraction decomposition of given rational */
def apart[E](frac: Rational[E]) = {
  implicit val ring: Ring[E] = frac.ring
  val facs = ring.factor(frac.denominator).map {case (f, exp) => f.pow(exp)}
  val (gcd,  nums) = solveDiophantine(facs.map(frac.denominator / _))
  val (ints, rats) = (nums zip facs)
    .map { case (num, den) => Rational(frac.numerator * num, den * gcd) }
    .flatMap(_.normal)       // extract integral parts from fractions
    .partition(_.isIntegral) // separate integrals and fractions
  rats :+ ints.foldLeft(Rational(ring(0)))(_ + _)
}

// partial fraction decomposition for rationals
// gives List(184/479, (-10)/13, 1/8, (-10)/47, 1)
val qFracs = apart( Q("1234213 / 2341352") )

// partial fraction decomposition for rational functions
val ufRing = Frac(UnivariateRingZp64(17, "x"))
// gives List(4/(16+x), 1/(10+x), 15/(1+x), (14*x)/(15+7*x+x^2))
val pFracs = apart( ufRing("1 / (3 - 3*x^2 - x^3 + x^5)") )
\end{lstlisting}
Function \texttt{apart[E]} is defined as a generic function which can be applied to fractions over elements of arbitrary rings (that should be Euclidean rings of course). Returning to our initial example where we've constructed a field $\GF(17, 3)[x, y, z]$, let us adjoin a new variable, say $W$, and construct the partial fraction decomposition in this complicated field:
\begin{lstlisting}[language=scala, numbers=left, firstnumber=80]
// partial fraction decomposition of rational functions
// in the ring %$\color{commcolor} Frac(\GF(17, 3)[x, y, z])[W]$%
implicit val uRing = UnivariateRing(ratRing, "W")
val W = uRing("W")
val fracs = apart(Rational(W + 1, (rx/ry + W.pow(2)) * (rz/rx + W.pow(3))))
\end{lstlisting}
The function call on the last line involves nearly all main components of \Rings library: from very basic algebra to multivariate factorization over sophisticated rings. 

The last example is actually closely related to a problem of ``partial fractioning" of loop integrals, i.e.\ transforming a loop-integral expression to a sum where each denominator contains only linearly (or, in general, algebraically) independent factors. The \texttt{apart} function implemented with just a few lines of code is actually almost ready to do this task in case of one-loop integrals (as applied in the last example, with few modifications, see e.g.\ \cite{Feng:2012iq}). However, for two- and more-loop integrals a more sophisticated code with \Groebner bases and Nullstellensatz certificates computation \cite{Meyer:2016slj,Leinartas} is required. Still, \Rings provides all the instruments to implement this task efficiently (in fact, there are built-in functions for computing Nullstellensatz certificates and partial fraction decompositions of multivariate fractions implemented in \href{http://static.javadoc.io/cc.redberry/rings/2.5/cc/redberry/rings/poly/multivar/GroebnerMethods.html}{GroebnerMethods} class).

The above examples show how powerful the features of \Rings in a combination with expressive and type-safe Scala syntax may be used to implement quite a non-trivial and generic functionality with little effort.

\section{Benchmarks}
\label{sec_bench}
The speed of \Rings was tested on four important problems: multivariate polynomial GCD (a key component of rational function arithmetic), multivariate polynomial factorization, univariate polynomial factorization, and \Groebner bases. The internal implementation of the corresponding algorithms uses nearly all components of the library, so it is a good indicator of the overall performance. 

All benchmarks presented below were executed on MacBook Pro (15-inch, 2017), 3,1 GHz Intel Core i7, 16 GB 2133 MHz LPDDR3 (unless otherwise noted). All benchmarking problems and instructions how to reproduce the results are available at \Rings \href{https://github.com/PoslavskySV/rings.benchmarks}{github page}. Performance of \Rings (v2.3.2) was compared to \textsc{Mathematica} (v11.1.1.0),  \textsc{Singular} (v4.1.0),  \textsc{FORM} (v4.2.0) \cite{Ruijl:2017dtg}, and \textsc{Fermat} (v6.19).

\subsection{Multivariate GCD and factorization}

Two corner cases are important for nearly all algorithms with multivariate polynomials: very dense problems and very sparse problems. In fact, most problems that arise in practice are sparse, especially those which arise in physics. On the other hand, dense problems can arise too, e.g.\ a very sparse polynomial $(x^{10} y^{10} z^{10} + 1)$ becomes quite dense under a simple substitution $x \to x + 1, y \to y + 1, z \to z + 1$. 

Another important fact is that many multivariate algorithms are quite sensitive to the properties of the ground ring. For example, the algorithms for GCD or factorization in $\ZZ[\vec X]$, $\ZZ_p[\vec X]$ with medium $p$ (32-bit) and $\ZZ_p[\vec X]$ with very small $p$ (e.g.\ $p = 2$) are quite different. 

Finally, both GCD and factorization may have two results: the trivial result (GCD of coprime or factorization of irreducible polynomial) and the non-trivial result. In most practical situations (especially in physics or calculus) the first case happens much more often than the second. So, it is a very important property of the algorithm to detect the trivial answer as early as possible.

In the below benchmarks we addressed the above considerations and performed separate tests for sparse and dense inputs over different ground rings with both non-trivial and trivial answers.

\subsubsection{\textbf{Multivariate GCD}}

To test the performance of GCD on sparse polynomials, we proceed as follows. Polynomials $a$, $b$, and $g$ of a fixed size were generated at random. Then the GCDs $gcd(ag, bg)$ (non-trivial, should result in multiple of $g$) and $gcd(ag+1, bg)$ (trivial) were calculated. The performance of GCD algorithms is quite sensitive to the shapes of the input polynomials, so two different methods for generating random polynomials were used. In both methods, monomials were generated independently, while the exponents in each monomial were distributed according to one of the two distributions:
\begin{itemize}
\item \textbf{uniform exponents distribution}: given two values $D_{\text{min}}$ and $D_{\text{max}}$, the exponent of each variable in the monomial is chosen independently and uniformly between $D_{\text{min}} \leq \text{exp} < D_{\text{max}}$, so the total degree of the monomial is obtained in the range $[N D_{\text{min}}, N D_{\text{max}}]$
\item \textbf{sharp exponents distribution}: given a single value $D_{\text{sum}}$, the exponents are chosen according to some multinomial distribution, so that the total degree is exactly $D_{\text{sum}}$. In particular, the exponent of some random first variable is chosen uniformly in $0 \leq \text{exp}_1 \leq D_{\text{sum}}$, then the exponent of some next variable is chosen uniformly in $0 \leq \text{exp}_2 \leq (D_{\text{sum}} - \text{exp}_1)$ and so on. Obviously, this method gives quite ``sharp'' distribution of exponents in the monomials.
\end{itemize}

\begin{figure}[h!]
\centerline{\includegraphics[width=0.8\textwidth]{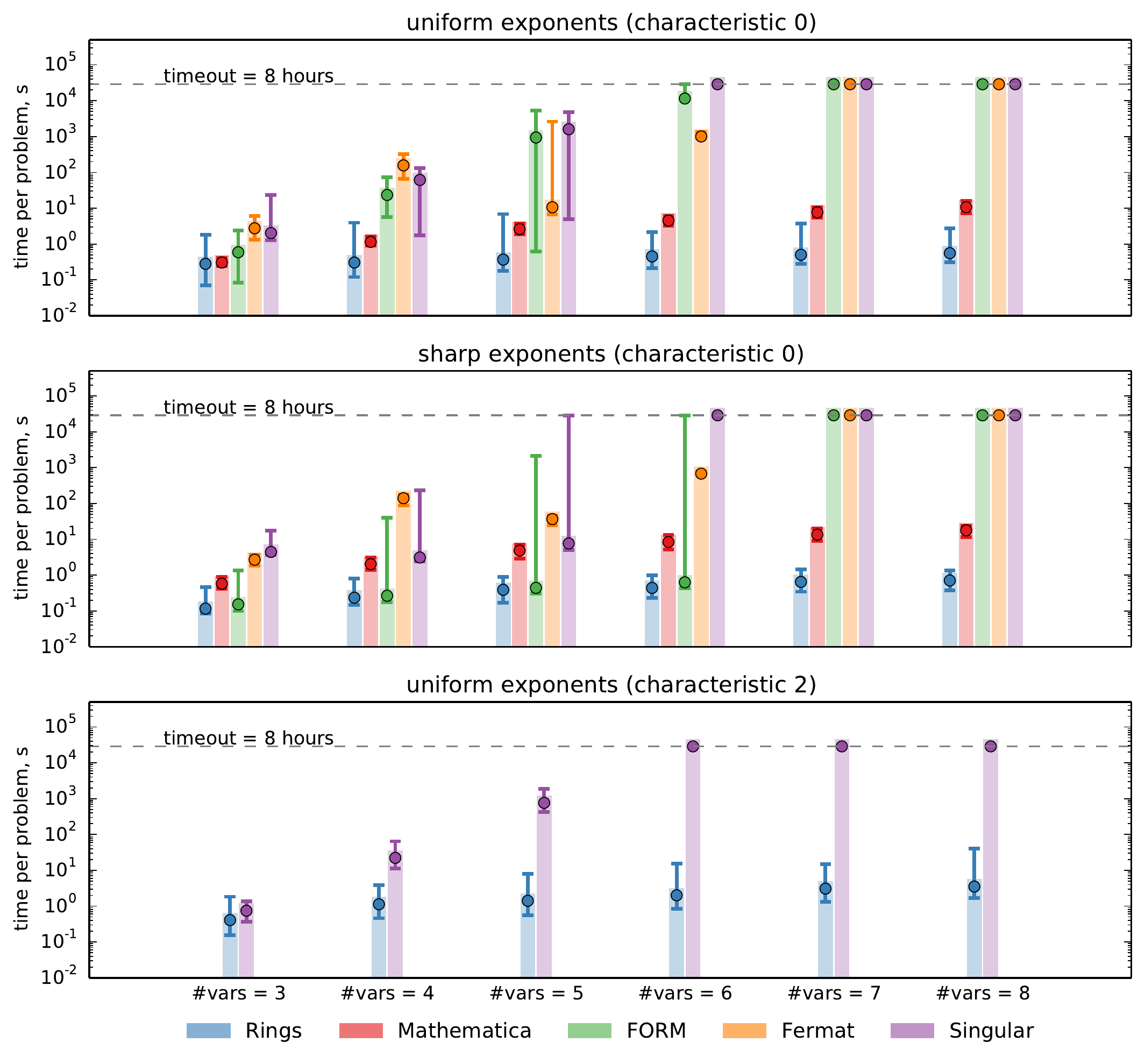} }
\caption{Dependence of multivariate GCD performance on the number of variables. Each problem set contains 110 problems, points correspond to the median times, and the error bands correspond to the smallest and largest execution time required to compute the GCD within the problem set. If computation of a single GCD took more than 8 hours (timeout) it was aborted and the timeout value was adjoined to the statistics. Two upper plots correspond to the two different methods of generating random polynomials (described in the text), both performed for characteristic zero (polynomials over $\ZZ$). The bottom plot for characteristic 2 (polynomials over $\ZZ_2$) was obtained using only the uniform method for the generation of random polynomials. The parameters of the random polynomials are given in the text.}
\label{fig_bench_gcd_nvars}
\end{figure}

Figure~\ref{fig_bench_gcd_nvars} shows how performance of different libraries behave with the increase of the number of variables. For this benchmark, polynomials $a$, $b$ and $g$ had size 40 (so the products $ag$ and $bg$ had at most 1600 terms each). For uniform distribution of exponents the values $D_{\text{min}} = 0$ and $D_{\text{max}} = 30$ were used. For sharp distribution of exponents $D_{\text{sum}} = 50$ was used. The two upper plots correspond to the characteristic zero (polynomials over $\ZZ$), and the lower plot to characteristic two (polynomials over $\ZZ_2$). In the latter case only \Rings and \Singular were compared since other libraries either don't support GCD in $\ZZ_p[\vec X]$ (\textsc{FORM} doesn't support and \textsc{Fermat} has support only for relatively large characteristic) or performance was extremely low (\Mathematica was nearly $10^3$ times slower in each of the considered problems). In all considered problems performance of \Rings was unmatched. Notably, its performance almost doesn't depend on the number of variables in such sparse problems.

\begin{figure}[h!]
\centerline{\includegraphics[width=\textwidth]{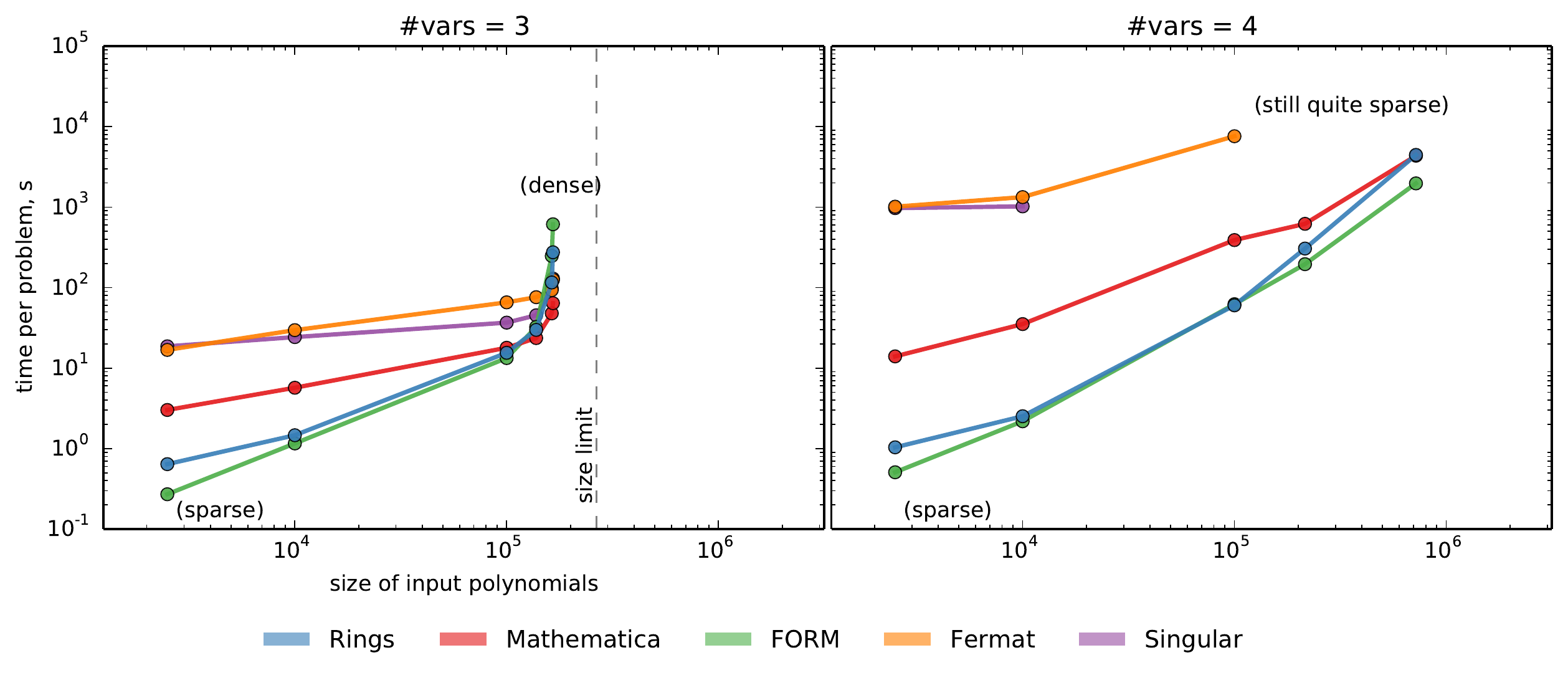} }
\caption{Dependence of multivariate GCD performance on the size of input polynomials. Polynomials over 3 variables were used for the left plot, and over 4 variables for the right. In the latter case \Singular and \textsc{Fermat} succeeded to obtain the answer within 24 hours only for a subset of considered problems. \Rings required nearly 15Gb of RAM for rightmost point on the right plot, while other tools required nearly 1.5 times less. }
\label{fig_bench_gcd_size}
\end{figure}

Figure~\ref{fig_bench_gcd_size} shows performance behavior with the increase of the input size. For this benchmark, polynomials $a$, $b$ and $g$ were generated using only sharp exponents distribution with $D_{\text{sum}} = 50$. Polynomials over 3 (left plot) and 4 (right plot) variables were used. Sizes of polynomials $a$, $b$ and $g$ were chosen from the set [50, 100, 500, 1000, 5000], so the products $ag$ and $bg$ had more than $10^6$ terms in the hardest case (those polynomials span more than 1 GB of space when saved in a text form). The left and right plots on Figure~\ref{fig_bench_gcd_size} actually correspond to a quite different cases.

Since there are only about $50^3$ different monomials of total degree 50 in 3 variables, the input polynomials $ag$ and $bg$ from the left plot of Figure~\ref{fig_bench_gcd_size} can have at most $2 \times 50^3$ different monomial terms, i.e.\ when their sizes tend to this limit, the input problem becomes very dense. This is reflected in the fact that the performance observably degrades for all considered tools. Only \textsc{Fermat} shows better performance in this case. In contrast, polynomials from the right plot of Figure~\ref{fig_bench_gcd_size} are still quite sparse even when their size is about $10^6$. In this (the hardest) case, \Rings required nearly 15 GB of RAM, while other tools required nearly 1.5 times less.

To further investigate the performance of GCD on dense problems, the following relatively dense problem was considered: 
\begin{eqnarray}
\label{eq_dense_gcd_a}
&& a = \left(1 + 3 x_1 + 5 x_2 + 7x_3 + 9x_4 + 11x_5 + 13x_6 + 15x_7\right)^7 - 1 \\
\label{eq_dense_gcd_b}
&& b = \left(1 - 3 x_1 - 5 x_2 - 7x_3 + 9x_4 - 11x_5 - 13x_6 + 15x_7\right)^7 + 1 \\
\label{eq_dense_gcd_g}
&& g = \left(1 + 3 x_1 + 5 x_2 + 7x_3 + 9x_4 + 11x_5 + 13x_6 - 15x_7\right)^7 + 3
\end{eqnarray}
The results are summarized in Table~\ref{tab_bench_gcd_dense}.

\begin{table}[h!]
\centering
\begin{tabular}{llrrrrr}
\toprule
 Problem &    Ground ring &     \Rings &  \Mathematica &     \textsc{FORM} &       \textsc{Fermat} &  \Singular \\
\midrule
  $gcd(ag, bg)\phantom{ + 1.}$ & $\ZZ$ & 104s &   115s &  148s &  1759s &    141s \\
  $gcd(ag, bg + 1)$  &  $\ZZ$ &   0.4s &     2s &    0.3s &     0.1s &      0.4s \\
  $gcd(ag, bg)\phantom{ + 1.}$ &  $\ZZ_{524287}$ &  25s &    33s &      N/A &     147s &     46s \\
  $gcd(ag, bg + 1)$  &  $\ZZ_{524287}$ &  0.5s &     2s &      N/A &     0.2s &      0.2s \\
\bottomrule
\end{tabular}
\caption{Time to compute $gcd(ag, bg)$ and $gcd(ag, bg + 1)$ for dense polynomials given in \eqref{eq_dense_gcd_a}, \eqref{eq_dense_gcd_b}, \eqref{eq_dense_gcd_g}. }
\label{tab_bench_gcd_dense}
\end{table}

\subsubsection{\textbf{Multivariate factorization}}

To test the performance of multivariate factorization on sparse polynomials we proceed as follows. Three polynomials $a$, $b$, and $c$ with 20 terms each were generated at random. Then the factorizations of $(abc)$ (non-trivial, should give at least three factors) and $(abc+1)$ (trivial, irreducible) were calculated. Uniform distribution of exponents with $D_{\text{min}} = 0$ and $D_{\text{max}} = 30$ (see above section) was used to generate random polynomials. For factorization benchmarks only \Rings and \Singular were used, since other tools either don't support multivariate factorization (\textsc{Fermat}), or has only very basic support (\Mathematica and \textsc{FORM}), so that nearly all considered sparse problems were intractable for them.

\begin{figure}[h!]
\centerline{\includegraphics[width=0.8\textwidth]{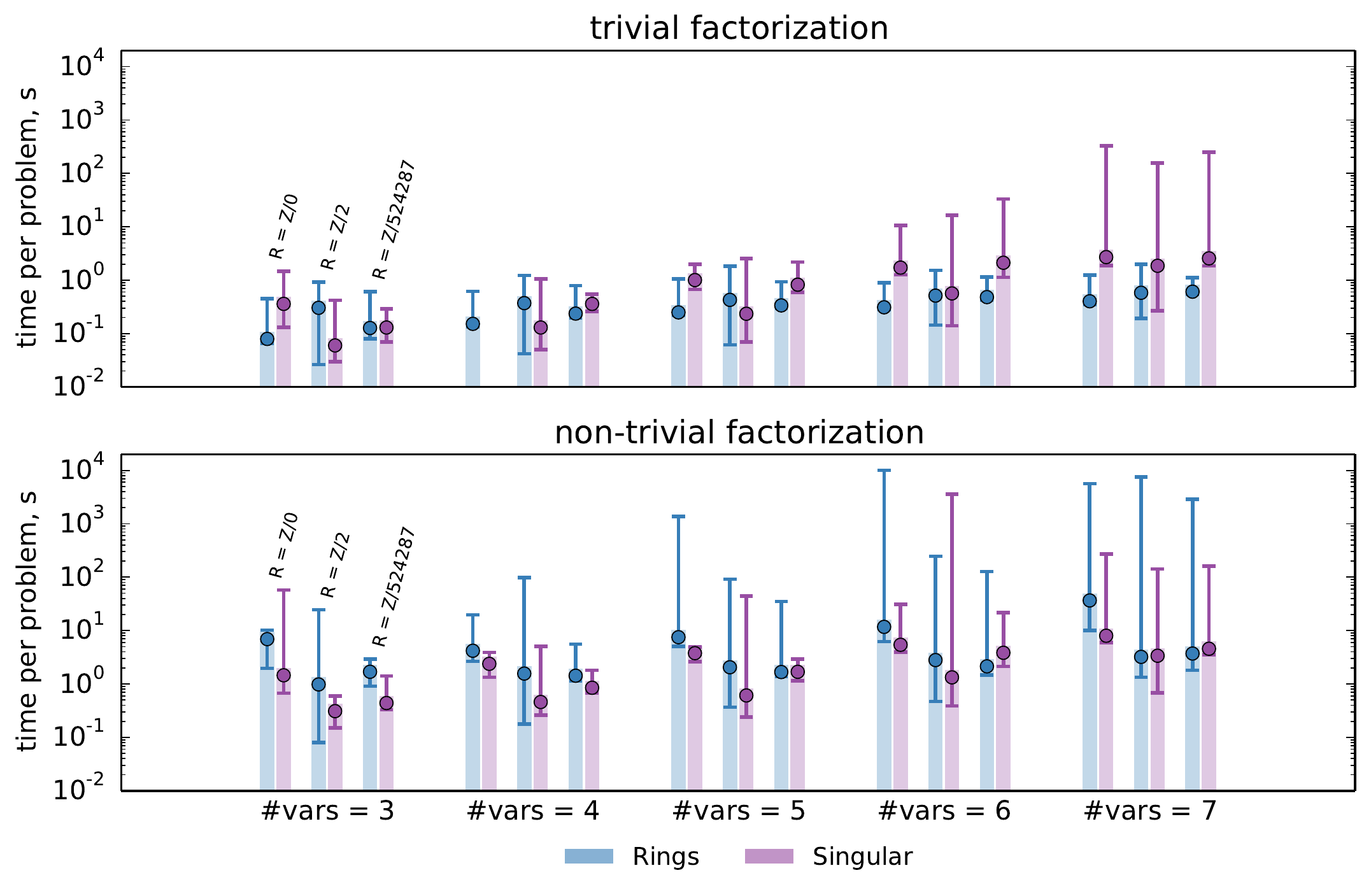} }
\caption{Dependence of multivariate factorization performance on the number of variables. Each problem set contained 110 problems, points correspond to the median times and the error bands correspond to the smallest and largest execution time required to compute the factorization within the problem set. The upper plot corresponds to the trivial factorization problems $(abc+1)$, the lower to non-trivial factorizations $(abc)$. Parameters of random polynomials are given in the text.}
\label{fig_bench_factor}
\end{figure}

\begin{figure}[h!]
\centerline{\includegraphics[width=0.8\textwidth]{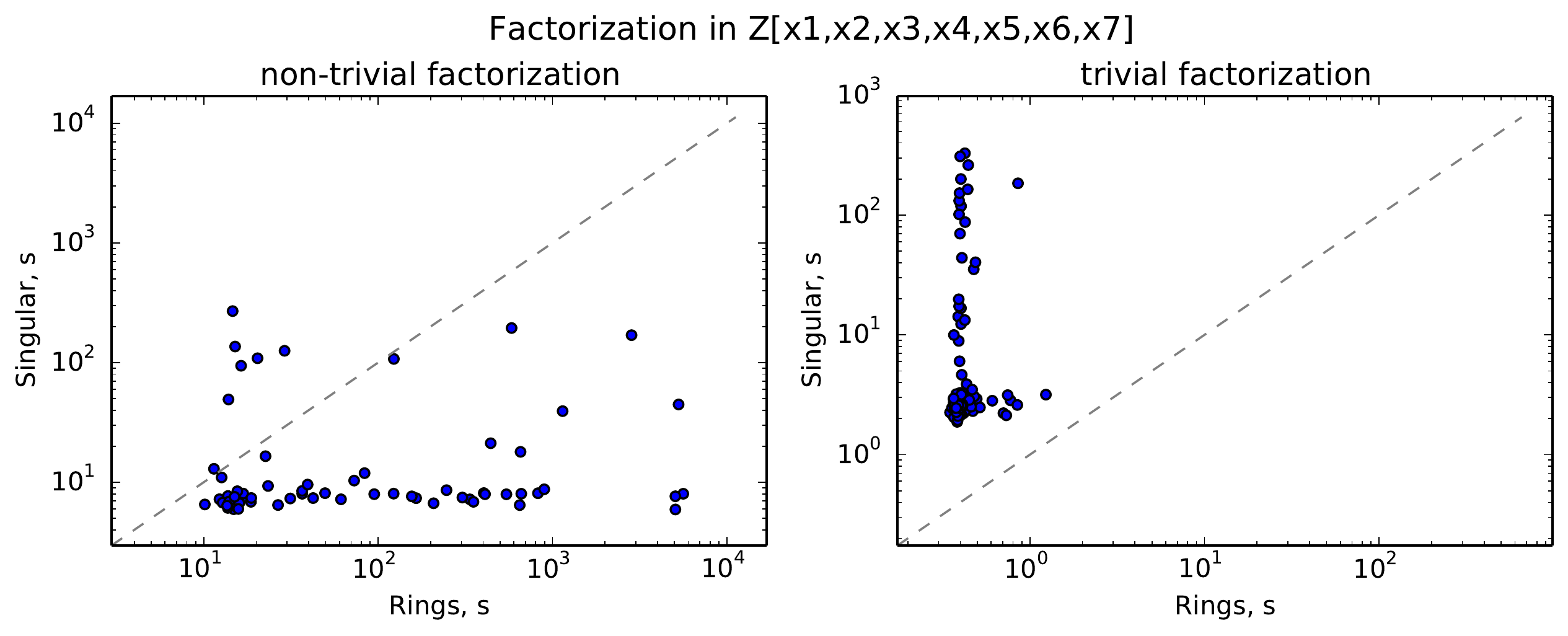} }
\caption{Comparison of \Rings and \Singular performance on multivariate factorization in $\ZZ[x_1, \dots, x_7]$}.
\label{fig_bench_factor_scattern}
\end{figure}

Figure~\ref{fig_bench_factor} shows how performance of multivariate factorization depends on the number of variables. It follows that the median time required to compute factorization changes quite slowly, while some outlying points (typically ten times slower than median values) appear when the number of variables becomes large. A more detailed analysis shows that those outliers are actually points where the sparse algorithms failed and the problem was solved using dense methods. Actually, modern multivariate factorization methods contain some heuristics to decide when to switch between sparse and dense algorithms, and these heuristics may be implemented in a quite different way in different libraries. Figure~\ref{fig_bench_factor_scattern} shows the typical time distributions for polynomial factorization of polynomials in seven variables over $\ZZ$ ground ring. Notably, \Singular crashed on some problems with either segfault or out-of-memory errors\footnote{bug report submitted to \Singular developers} --- such points were excluded from comparison. 

\begin{table}[h!]
\centering
\begin{tabular}{llrrrr}
\toprule
Problem & Ground ring  &   \Rings & \Singular & \Mathematica \\
\midrule
 $factor(p_1)$  & $\ZZ_{\phantom{2}}$  & 55s    & 20s        & 271s \\
 $factor(p_1)$  & $\ZZ_2$                      & 0.25s & $>$ 1h & N/A \\
 $factor(p_1)$  & $\ZZ_{524287}$          & 28s    & 109s    & N/A \\
 $factor(p_2)$  & $\ZZ_{\phantom{2}}$   & 23s    & 12s        & 206s \\
 $factor(p_2)$  & $\ZZ_2$           &  6s & 3s & N/A \\
 $factor(p_2)$  & $\ZZ_{524287}$  &26s    & 9s    & N/A \\
\bottomrule
\end{tabular}
\caption{Time to compute factorization of dense polynomials given in \eqref{eq_dense_factor_1} and \eqref{eq_dense_factor_2}.}
\label{tab_bench_factor_dense}
\end{table}

To further investigate performance on dense problems, factorization of the following relatively dense polynomials was tested:
\begin{eqnarray}
\label{eq_dense_factor_1}
&&p_1 = (1 + 3 x_1 + 5 x_2 + 7 x_3 + 9 x_4 + 11 x_5 + 13 x_6 + 15 x_7)^{15} - 1\\
&&\label{eq_dense_factor_2}\begin{split}p_2 = -1 + (1 + 3x_1x_2 + 5x_2x_3 + 7x_3x_4 + 9x_4x_5 + 11x_5x_6 + 13x_6x_7 + 15x_7x_1)^3\\
       \quad \times (1 + 3x_1x_3 + 5x_2x_4 + 7x_3x_5 + 9x_6x_5 + 11x_7x_6 + 13x_6x_1 + 15x_7x_2)^3\\
        \quad \quad \times (1 + 3x_1x_4 + 5x_2x_5 + 7x_3x_6 + 9x_6x_7 + 11x_7x_1 + 13x_6x_2 + 15x_7x_3)^3\end{split}
\end{eqnarray}
It is worth to admit that the first polynomial has unit leading coefficient with respect to any of its variables, while the second has not (these two corner cases are very differently treated by many known factorization algorithms). The results are listed in Table~\ref{tab_bench_factor_dense}. It is seen that \Rings and \Singular have comparable performance.

\subsection{\bfseries \Groebner bases}
Performance of \Groebner bases was tested on classical Katsura and cyclic polynomial systems. In all cases graded reverse lexicographic order was used to compute the \Groebner basis. Results are summarized in Table~\ref{ref_tab_groebner}. Timings are in general comparable between \Rings and \Singular for polynomial ideals over $\ZZ_p$ while for $\QQ$ \Rings behaves worse. It should be noted that for very hard problems a much more efficient dedicated tools like \textsc{FGB} \cite{fgb} (proprietary) or \textsc{OpenF4} \cite{openF4} (open source) exist.

\begin{table}[h!]
\begin{center}
\begin{tabular}{llrrr}
\toprule
Problem &  Ground ring &     \textsc{Rings} & \textsc{Mathematica} &  \textsc{Singular}\\
\midrule
cyclic-7 & $\ZZ_{1000003} $ &       3s &  26s &      N/A    \\
cyclic-8 & $\ZZ_{1000003} $ &      51s & 897s &     39s       \\
cyclic-9 & $\ZZ_{1000003} $ &   14603s & $\infty$ &   8523s  \\
\midrule
katsura-7 & $\ZZ_{1000003} $ &   0.5s & 2.4s &      0.1s \\
katsura-8 & $\ZZ_{1000003} $ &         2s & 24s &      1s \\
katsura-9 & $\ZZ_{1000003} $ &               2s & 22s &      1s  \\
katsura-10 & $\ZZ_{1000003} $ &              9s & 216s &      9s \\
katsura-11 & $\ZZ_{1000003} $ &            54s &   2295s &   65s \\
katsura-12 & $\ZZ_{1000003} $ &          363s &  28234s &   677s \\
\midrule
katsura-7 & $\QQ$ &              5s &  4s &     1.2s \\
katsura-8 & $\QQ$ &              39s &  27s &     10s \\
katsura-9 & $\QQ$ &              40s &  29s &     10s \\
katsura-10 & $\QQ$ &              1045s &  251s &     124s \\
\bottomrule
\end{tabular}
\caption{Time required to compute \Groebner basis in graded reverse lexicographic order. For cyclic-9 problem \Rings required about 50 GB of RAM to succeed, so it was executed on Intel(R) Xeon(R) CPU E5-2683 v3 @ 2.00GHz, 512 GB 2133 MHz DIMM. \label{ref_tab_groebner}}
\end{center}
\end{table}

\subsection{\bfseries Univariate factorization in finite fields} 
Efficient univariate polynomial arithmetic is crucial for applications in number theory and cryptography, not to mention the fact that most multivariate algorithms use univariate in the basis. Univariate factorization in finite fields actually covers almost all low-level aspects of univariate arithmetic, so it is a quite good indicator of its overall performance.

\begin{figure}[h!]
\centerline{\includegraphics[width=0.95\textwidth]{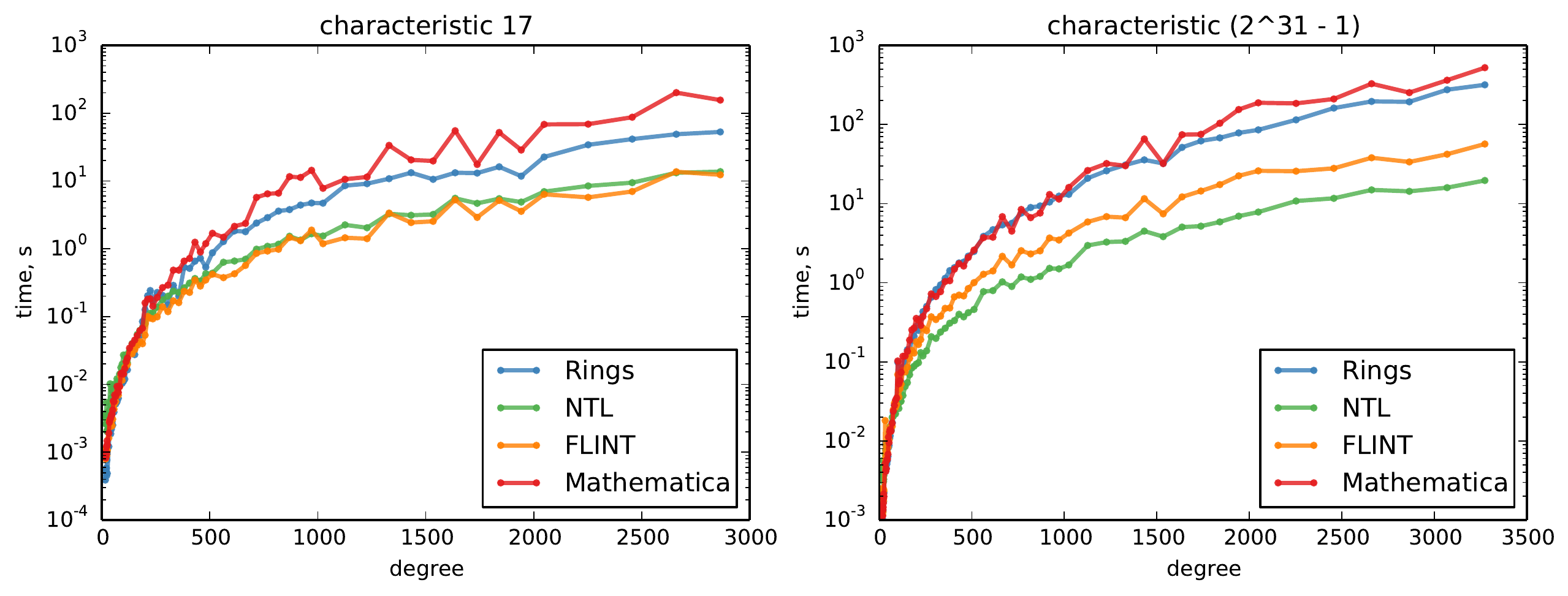}}
\caption{Univariate factorization performance on polynomials \eqref{eq_uni} over $\ZZ_{17}$ (left) and $\ZZ_{2^{31}-1}$ (right). } 
\label{fig_uni_fac}
\end{figure}

To test the univariate factorization performance we used the following polynomials:
\begin{equation}
\label{eq_uni}
p_{\mbox{\scriptsize deg}}[x] = 1 + \sum_{i = 1}^{ i \leq \mbox{\scriptsize deg}} i \times x^i
\end{equation}
and measured time to factor over $\ZZ_{17}$ and $\ZZ_{2^{31}-1}$. The speed was compared to \textsc{NTL} (v10.4.0) and \textsc{FLINT} (v2.5.2). Results are shown in Figure~\ref{fig_uni_fac}. It is seen that while \Rings has better performance than \Mathematica, it has worse asymptotic at large degrees than \textsc{NTL} and \textsc{FLINT}. Since \Rings uses the same algorithms for univariate polynomial factorization as \textsc{NTL} and \textsc{FLINT} (see next Section), the difference comes from the fact that \textsc{NTL}/\textsc{FLINT} use algorithms with better asymptotic complexity for univariate polynomial multiplication (namely algorithms based on Discrete Fast Fourier Transformation).

\section{Implementation notes}
\label{sec_impl}
\Rings is separated into several main components. They include:
\begin{itemize}
\item \texttt{rings.bigint}: arbitrary precision arithmetic
\item \texttt{rings.primes}: prime numbers, prime factorization and primality test
\item \texttt{rings.poly.univar}: univariate polynomials and related algorithms
\item \texttt{rings.poly.multvar}: multivariate polynomials and related algorithms
\item \texttt{rings.scaladsl}: type classes and syntax definitions for Scala API
\end{itemize}
In the subsequent sections we briefly discuss each component, but before some comments about the general design of the library are in order.

\subsection{General design}

\Rings is written in object-oriented paradigm with a deep use of abstraction, inheritance, and generic programming. On top of the Java API the Scala API brings powerful strongly typed functional programming model, which allows to write short and expressive code for applications.

The central top-level interface \texttt{Ring[E]} defines ring operations on elements of type \texttt{E} (addition, multiplication, etc.) and ring properties (cardinality, characteristic, etc.). All built-in rings in the library are subtypes of \texttt{Ring[E]}. For example, \texttt{IPolynomialRing[Poly, Coef]} is a subtype which is a super type for all polynomial rings. The second central interface is \texttt{IPolynomial[Poly]}, which is the top-level interface for all polynomial types (\Rings uses several different implementations of polynomials, discussed in the subsections below). The following example illustrates generic programming with abstract mathematical types in \Rings with the use of functional programming in Scala:
\begin{lstlisting}[language=scala]
/** Lagrange univariate polynomial interpolation: %$\color{commcolor}p(x) = \sum_i p(x_i) \Pi_{j \ne i} \frac{x_{\phantom{i}} - x_j}{x_i -x_j}$% */
def interpolate[Poly <: IUnivariatePolynomial[Poly], Coef]
(points: Seq[Coef], values: Seq[Coef])
(implicit ring: IUnivariateRing[Poly, Coef]) : Poly = {
  // implicit coefficient ring (setups algebraic operators on type Coef)
  implicit val cfRing: Ring[Coef] = ring.cfRing
  if (!cfRing.isField) throw new IllegalArgumentException
  points.indices
    .foldLeft( ring(0) ) { case (sum, i) =>
      sum + points.indices
        .filter(_ != i)
        .foldLeft( ring(values(i)) ) { case (product, j) =>
          product * (ring.`x` - points(j)) / (points(i) - points(j))
        }
    }
}
\end{lstlisting}
Function \texttt{interpolate} from the above example is a generic function which applies to a sequence of interpolation \texttt{points} and polynomial \texttt{values} at these \texttt{points}. Note that all types of objects (which are omitted in code) are inferred at compile time. This powerful feature allows to use two distinct implicit rings in the function body: \texttt{ring} which defines algebraic operations on type \texttt{Poly} (polynomials) and \texttt{cfRing} which defines algebraic operations on type \texttt{Coef} (polynomial coefficients). Function \texttt{interpolate(points, values)} can be applied in any ring of univariate polynomials:
\begin{lstlisting}[language=scala]
// coefficient field %$\color{commcolor}Frac(\,\ZZ_{2}[a,b,c]\,)$%
val cfRing = Frac(MultivariateRingZp64(2, Array("a", "b", "c")))
val (a, b, c) = cfRing("a", "b", "c")
// univariate ring %$\color{commcolor}Frac(\,\ZZ_{2}[a,b,c]\,)[x]$%
implicit val ring = UnivariateRing(cfRing, "x")
// interpolate with Lagrange formula
val data = Seq(a -> b, b -> c, c -> a)
val poly = interpolate(data.map(_._1), data.map(_._2))
assert(data.forall { case (p, v) => poly.eval(p) == v })
\end{lstlisting}
In this example, the compile-time type of \texttt{poly} (which is again omitted for shortness) will be automatically inferred as \texttt{UnivariatePolynomial[Rational[MultivariatePolynomialZp64]]}.

\subsection{Arithmetic with integers}
Java has out-of-the-box support for arbitrary-precision arithmetic with integer and floating-point numbers via the \texttt{BigInteger} and \texttt{BigDecimal} classes. Historically, Java's built-in implementation was much slower than, for example, the well-known GNU Multiple Precision Arithmetic Library (GMP) \cite{Granlund12}, which has became the de facto standard of arbitrary-precision software. The difference was really noticeable in the real-world computing applications. The situation became much better with the release of Java 8, which contained Karatsuba and Toom--Cook \cite{Knuth} methods for integer multiplication. \Rings uses a further improved implementation of \texttt{BigInteger}\footnote{a fork of \href{https://github.com/tbuktu/bigint}{https://github.com/tbuktu/bigint}} with the support of Sch\"onhage--Strassen \cite{Knuth} multiplication and Barrett \cite{Barrett1987} division for very large integers. With these improvements, performance of arbitrary-precision arithmetic has ceased to be the bottleneck and now has a little impact on a high-level methods like polynomial GCD or factorization (although it is still not so performant as in GMP, and still may be the bottleneck for computing some challenging \Groebner bases over $\QQ$).

Another important implementation aspect concerns arithmetic in the ring $\ZZ_p$ with $p < 2^{64}$, that is integer arithmetic modulo some machine number. Though it may be hidden from the user's eye, arithmetic in this ring actually lies in the basis of the most part of fundamental algorithms and directly affects performance of nearly all computations. On the CPU level the modulo operation is implemented via \texttt{DIV} instruction (integer division), which is known to be very slow: for example on the recent Intel Skylake architecture \texttt{DIV} has 20--80 times worse throughput than \texttt{MUL} instruction \cite{agner}. Hopefully, arithmetic operations in $\ZZ_p$ are done modulo a fixed modulus $p$, which allows to make some preconditioning on $p$ and reduce \texttt{DIV} operations to \texttt{MUL}. The idea is the following \cite{Barrett1987}: given a fixed $p$ we compute once the value of $magic = [2^{n} / p]$ with a sufficiently large $n$ (so that $magic$ is some non-zero machine number), and then for arbitrary integer $a$ we have $[a / p] = (a \times magic) / 2^n$, so the \texttt{DIV} instruction is replaced with one \texttt{MUL} and one \texttt{SHIFT} (division by a power of two is just a bitwise shift, very fast). The actual implementation in fact requires some more work to do (for details see Chapter 10 in \cite{HackersDelight}). The implementation used in \Rings gained 3-fold increase in speed for modular arithmetic in comparison to the implementation via native CPU instructions\footnote{see \url{https://github.com/PoslavskySV/libdivide4j} which is a Java port of well known C/C++ \texttt{libdivide} library (\url{https://libdivide.com})}.

\subsection{Prime numbers}
Problems with prime numbers (primality test, integer factorization, generating primes, etc.) are key computational problems of modern number theory and cryptography. Though \Rings doesn't address these problems rigorously, it contains methods that are required for efficient implementation of some high-level algorithms with polynomials. These methods include:
\begin{itemize}
\item \texttt{SmallPrimes} --- factorization and primality test for integers less than $2^{32}$. It uses the Miller--Rabin \cite{Miller,Rabin} probabilistic primality test in such a way that result is always guaranteed\footnote{code is adapted from Apache Commons Math library \href{http://commons.apache.org/proper/commons-math/}{http://commons.apache.org/proper/commons-math/}} and trial divisions for integer factorization (which is very fast for such small inputs).
\item \texttt{BigPrimes} --- factorization and primality test for arbitrarily large integers. It switches between Pollard-$\rho$ \cite{Pollard1975}, Pollard-P1 \cite{pollard_1974} and Quadratic Sieve \cite{qsieve,crandall2006prime} algorithms for prime factorization and uses Miller--Rabin probabilistic and Lucas \cite{crandall2006prime} strong primality tests.
\item \texttt{SieveOfAtkin} and \texttt{PrimesIterator} --- simple implementation of the sieve of Atkin \cite{atkin} and methods for primes generation.
\end{itemize}

\begin{figure}
\includegraphics[width=\textwidth]{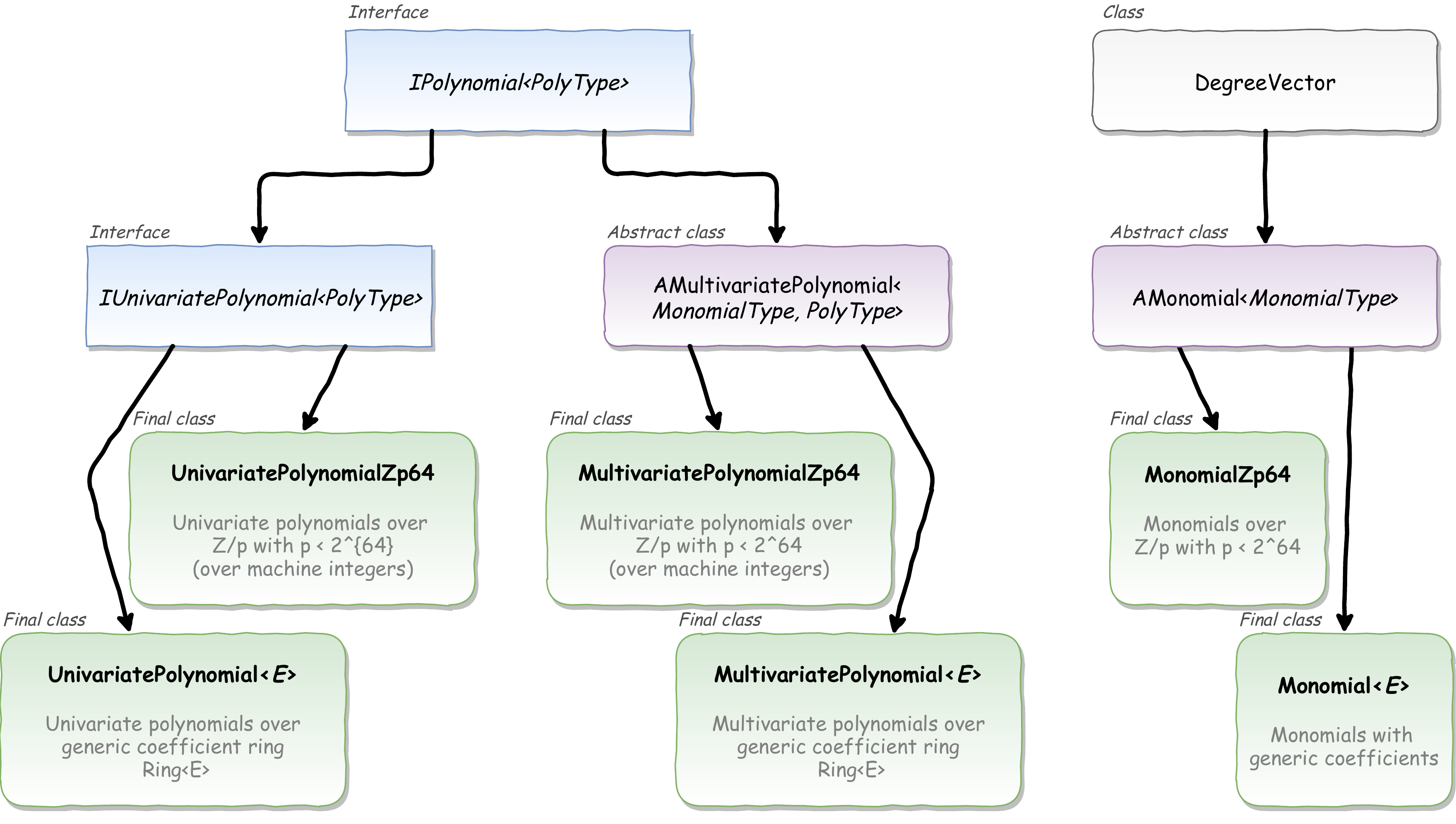}
\caption{Inheritance tree of polynomial types.}
\label{fig_uml}
\end{figure}

\subsection{Univariate polynomials}
\label{sec_uni_poly}
Univariate polynomials are implemented as dense arrays of coefficients. There are two distinct implementations (see Fig.~\ref{fig_uml}): polynomials over generic ground rings (\texttt{UnivariatePolynomial[E]}) and polynomials over $\ZZ_p$ with $p < 2^{64}$ (\texttt{UnivariatePolynomialZp64}). The latter one uses primitive long integers to store coefficient data, that's why it is implemented separately: to avoid the considerable cost of Java's boxing and unboxing.

Multiplication of polynomials --- the most used and time consuming primitive operation  --- is handled via Karatsuba's method, which reduces the cost of single multiplication from $O(n^2)$ to $O(n^{\log_2 3})$. The counterpart of multiplication -- polynomial division with remainder --- switches between classical division and division via Newton's iteration (see Sec.~9 in \cite{CAHB}). The Newton iteration is essentially similar to the trick \Rings uses for fast arithmetic in $\ZZ_p$: it allows to reduce the division to multiplication by a some precomputed ``inverse'' polynomial followed by some trivial coefficients rearrangement. This techniques becomes especially useful when several divisions by a fixed polynomial are required (case arising in many algorithms): one can compute the inverse only once and use it in all further divisions. This is also used in the implementation of Galois fields $\GF(p, q)$ which are represented as univariate polynomials over $\ZZ_p$ reduced modulo some minimal polynomial $r(x)$ with $\mbox{deg}(r) = q$ (that is $\ZZ_p[x]/\langle r(x) \rangle$). Fast division is easily accessible via high-level interface:
\begin{lstlisting}[language=scala]
// some univariate polynomials
val divider = ... ; val dividend = ...
// quotient and remainder using classical division methods
val (divPlain, remPlain) = dividend /%\%% divider
// precomputed Newton's inverses
implicit val inverse = divider.precomputedInverses
// quotient and remainder using Newton's iteration (fast)
val (divFast, remFast) = dividend /%\%\%% divider

assert((divPlain, remPlain) == (divFast, remFast))
\end{lstlisting}

Univariate GCD is implemented with the use of several different methods. For polynomials over finite fields it switches between the classic Euclid algorithm (for small polynomials) and fast Half-GCD algorithm (for medium and large polynomials, see Sec.~11 in \cite{CAHB}). For polynomials over $\ZZ$ and $\QQ$, the small-primes version of Brown's modular GCD algorithm \cite{Brown} is used. Similar modular algorithms \cite{LaMc89,Enca95} are used for GCDs over algebraic number fields $\QQ(\alpha)$. In all other cases, the subresultant GCD algorithm is used.

Univariate polynomial factorization is implemented with the use of Cantor--Zassenhaus approach. Factorization over finite fields is done either via plain Cantor--Zassenhaus algorithm \cite{CanZas} (for medium-sized polynomials) or Shoup's baby-step giant-step algorithm \cite{Shou95} (for large polynomials or for finite fields with large characteristic). Factorization of polynomials over $\ZZ$ is done first by factoring modulo some 32-bit prime, followed by a true factor reconstruction via Hensel lifting. Reconstruction of true factors from the modular ones is what dominates in the timing of factorization over $\ZZ$. \Rings uses a mixed variant of linear and quadratic Hensel lifting (see Sec.~6 in \cite{Geddes}) in order to lift modular factors up to sufficiently large modulus. The true factors are then finally reconstructed by a brute-force enumeration of all possible factor combinations (``naive'' recombination). This technique is quite fast for practical applications: in randomized tests most time is spent in actual Hensel lifting, while final recombination takes much less. Nevertheless, the bad case has exponential complexity. It takes place on a special class of polynomials --- Swinnerton-Dyer polynomials \cite{Zippel93} --- which have nontrivial factorization over $\ZZ_p$ (for any $p$) but irreducible over $\ZZ$. On these inputs the algorithm degrades to enumeration of all possible modular factor combinations which are $O(2^r)$, $r$ being the number of modular factors. The drawback is fixed by van Hoeij's algorithm \cite{Hoej02} which uses LLL lattice reduction \cite{LLL82} to reconstruct true factors in $O(r^3)$, but this algorithm is not implemented in \Rings yet. For factorization of polynomials over algebraic number fields $\QQ(\alpha)$ \Rings uses Trager's algorithm \cite{Trag76}.

\subsection{Multivariate polynomials}
\Rings uses sparse distributed representation of multivariate polynomials. The underlying data structure is a red--black map of degree vectors to monomials: Java's built-in \texttt{TreeMap[DegreeVector, Monomial]} is used. As in the case of univariate polynomials, there are two distinct implementations (see Fig.~\ref{fig_uml}): polynomials over generic ground rings (\texttt{MultivariatePolynomial[E]}) and polynomials over $\ZZ_p$ with $p < 2^{64}$ (\texttt{MultivariatePolynomialZp64}).

Multiplication of multivariate polynomials is done via Kronecker substitution (see e.g.\ Sec.~8.4 in \cite{CAHB} and \cite{Monagan2011Mul}), i.e.\ by reducing multivariate multiplication to a multiplication of sparse univariate polynomials by appropriate replacement of variables. Sparse univariate multiplication is done with a plain $O(nm \log(nm))$ algorithm (where $n$ and $m$ are numbers of nonzero terms in polynomials and the logarithm occurs due to search in the tree), hence the complexity of multivariate multiplication is also $O(nm \log(nm))$. Nevertheless, Kronecker substitution allows to significantly (up to an order of magnitude) reduce the constant overhead: \texttt{TreeMap[DegreeVector, Monomial]} (sparse multivariate) is replaced with \texttt{TreeMap[Long, Monomial]} (sparse univariate) which is significantly faster (both on \texttt{get()} and \texttt{put()} operations) since comparing of \texttt{long} keys is much faster than comparing of \texttt{DegreeVector}s \cite{Monagan2011Mul}. Actually, one can speed up even more by using \texttt{HashMap} instead of \texttt{TreeMap} since order of the intermediate terms is irrelevant. The strategy described above performs especially well on sparse polynomials (which is the common use case). It is also satisfactory for a dense ones, while certainly in this case it would be better to switch to dense recursive representation and use asymptotically fast univariate multiplication.

Multivariate GCD switches between Zippel-like sparse interpolation algorithms \cite{Zipp79,dKMW05} and Enhanced Extended Zassenhaus algorithm (EEZ-GCD) \cite{Wang80}. The latter is used only on a very dense problems (which occur rarely), while the former is actually used in most cases. While sparse algorithms are designed to perform well on sparse polynomials, they are actually quite fast on most dense problems too. These algorithms require that the ground ring contains a sufficient number of elements (so they will always fail in e.g.\ $\ZZ_2[\vec X]$). When the cardinality of a ground ring is not sufficiently large, \Rings switches to a Kaltofen--Monagan generic modular algorithm \cite{KalM99}, which performs very well in practice. We have also implemented Brown's \cite{Brown} and Extended Zassenhaus (EZ-GCD) \cite{MosY73} algorithms, but they are considerably slower both in asymptotic complexity and in practice. For polynomials over algebraic number fields, the modular approach with either sparse (Zippel) or dense (EEZ-GCD) interpolation is used with further rational-number reconstruction \cite{Enca95}. It is worth to mention that the implementation of multivariate GCD in \Rings is thoroughly optimized and takes more than 6,000 lines of code: all corner cases are carefully examined and specifically optimized and a lot of heuristics is used, based on thousands of benchmarks with both randomized and special input. As result the algorithms actually implemented in \Rings are quite different from those described in the original papers and the similarity remains only in the basic ideas.

It is worth briefly sketching several heuristic tricks used in \Rings to boost multivariate GCD algorithms. The actual amount of calculations in any algorithm is strictly dependent on the GCD degree bounds (these bounds are the input parameter in almost all algorithms). The obvious upper bound, which is usually used in practice, is the minimal degree of input polynomials $n_i^{\text{(est.)}} = \min(n_i^{(a)}, n_i^{(b)})$. This estimation is too rough since GCD typically has smaller degrees (or even trivial). For example, the amount of work performed in Brown's algorithm is proportional to $S_B = \Pi_{i = 2}^N n^{(\text{est.})}_i$ where $n_i$ is the true degree of $i$-th variable in the GCD and $n_i^{(\text{est.})}$ is the estimated degree bound; for Zippel's algorithm it is $S_Z = \Sigma_{i = 1}^N n^{(\text{est.})}_i n_i^2$ (monic case). The overestimation of GCD degree bounds by a factor of 2 thus leads to a $2^{(N-1)}$-fold increase in the amount of performed calculations for Brown's and 2-fold for Zippel's algorithms. All these extra calculations are redundant. To get rid of them (and thereby significantly speed up the average case) \Rings tries to substitute some values for all variables except the $i$-th one and computes the univariate GCD (very cheap). Doing this for each $i$, a very strict GCD degree bounds are obtained (and a trivial GCD is discovered already at this step in most cases). For polynomials over $\ZZ$, evaluation at random integer points may be already quite expensive (very large numbers may appear), so in this case univariate GCDs are evaluated modulo some prime number, which still gives quite strict bounds in practice.

Obtaining strict degree bounds may be also very helpful in some special cases. For example, it is not rare that the estimated degree bound for some $i$-th variable is zero (while this variable may still appear in both input polynomials). In this case one can compute GCD $gcd(a, b) \in R[x_1, \dots, x_N]$ as $gcd(\{ \tilde a_j \}, \{ \tilde b_j \} )$ where $\{ \tilde a_j \}$ and $\{ \tilde b_j \}$ are sets of coefficients of $a$ and $b$ considered as polynomials in $R[x_1, \dots, x_{i-1}, x_{i+1}, \dots, x_N][x_i]$. These coefficients are smaller and have $(N-1)$ variables, so pairwise GCDs are much faster. Moreover, the GCD of several polynomials is in fact as fast as the GCD of a single pair: one can take e.g.\ the first polynomial in sequence and compute its GCD with the sum of all other polynomials --- in most cases this will be the correct answer (for details see Section 6.9 of \cite{CAHB}).

Another important trick concerns representation of polynomials and fast evaluation. For sparse problems and sparse algorithms (e.g.\ Zippel-like) the sparse distributed representation of polynomials is very beneficial. However, evaluation of huge polynomials (with millions of terms) in sparse representation is very slow. This slow evaluation may become a real bottleneck in some GCD algorithms. To overcome this, in case of huge inputs \Rings may switch to a sparse recursive representation and apply fast Horner rule for polynomial evaluation. This trick has especially high performance impact in case of Zippel-like algorithms, where several evaluations of the same polynomial but with different value substitutions performed: one can compute sparse recursive representation once (computationally expensive) and then use it every time for evaluation (very fast).

The main idea of most multivariate factorization algorithms is simple: substitute some values for all variables but one, then perform univariate factorization and finally reconstruct multivariate factors with the use of Hensel lifting (lift from univariate factors to multivariate). The differences between algorithms lies in the details. In particular, the common problem, which is addressed differently by different algorithms, is that for efficient Hensel lifting it is required to precompute somehow the leading coefficients of true multivariate factors. \Rings uses the idea proposed by Kaltofen \cite{Kalt85}, with major modifications due to Lee \cite{LeeM13}. First, the multivariate case is reduced to bivariate by substituting all but two variables with some random values (actually, \Rings does several substitutions for different sets of variables to obtain more information about factorization pattern). For factoring bivariate polynomials, the very efficient Bernardin algorithm \cite{BerM97, Bern99} is used. Additionally, \Rings performs some fast early checks based on Newton polygons to ensure that there is a nontrivial factorization \cite{GAO2001501}. Then the leading coefficients of true multivariate factors are precomputed based on the square-free decomposition of the leading coefficient of the initial polynomial (while in the original Kaltofen paper the distinct-variable decomposition was used). The final Hensel lifting is then done via a Zippel-like sparse method: instead of using variable-by-variable dense ideal-adic interpolation, the problem of lifting is reduced to a system of (in general non-linear) equations which may be solved efficiently in many cases. If it is not possible to solve the system with the available methods, \Rings switches to the standard ideal-adic Hensel lifting algorithm. The latter one is also optimized by using the quasi-dense approach, which is a generalization of Bernardin's bivariate algorithm \cite{BerM97, Bern99, LeeM13}. As a result, the final algorithm for multivariate polynomial factorization implemented in \Rings is quite efficient on both sparse and dense inputs. Similar to the case of univariate factorization, there is one special bad case --- bivariate Swinnerton-Dyer \cite{Zippel93} polynomials --- which have nontrivial univariate factorization (for any value substituted for second variable) but still are irreducible. For these polynomials (which are impossible to imagine occurring in practice) the \Rings algorithm will degrade to $O(2^r)$ complexity, $r$ being the number of univariate factors. Again, the drawback may be fixed by the analog of van Hoeij's algorithm \cite{Hoej02, LeeM13} which uses LLL lattice reduction \cite{LLL82}, but it is not implemented in \Rings yet. For factorization of polynomials over algebraic number fields $\QQ(\alpha)$ \Rings uses Trager's algorithm \cite{Trag76}.

Development of new efficient algorithms for computing \Groebner bases is a very active research area, and many new algorithms are proposed every year. \Rings uses different algorithms for computing \Groebner bases of polynomial ideals, depending on monomial order and coefficient ring used. In all algorithms the Gebauer--Moller installation \cite{GebM88, BecW93} of Buchberger criteria is applied. \Groebner bases over finite fields for graded orders are computed with the use of Faugere's F4 algorithm \cite{Faug99} with fast sparse linear algebra \cite{FauL10} and simplification algorithm due to \cite{JouV11}. \Groebner bases for non-graded orders are first computed with respect to some graded order and then the ``change of ordering algorithm'' is applied. The latter one is based on Hilbert-driven methods \cite{Trav96} (with optional use of homogenization--dehomogenization steps). \Groebner bases over $\QQ$ may either use F4 or Buchberger algorithm \cite{Buch76} directly or, in some cases, switch to modular algorithm \cite{Arno03} (especially for small number of indeterminates). If the Hilbert--Poincare series of ideal is known in advance, the Hilbert-driven algorithm \cite{Trav96,CLOS97} will be used. If non of the above cases apply, the plain Buchberger algorithm \cite{Buch76,BecW93,CLOS97} is used. The latter may use either normal syzygy selection strategy (for graded orders) or sugar strategy (e.g. for lexicographic order) \cite{GMNR88}.

\section{Conclusion and future work}
\Rings is a high-performance and lightweight library for commutative algebra that provides both basic methods for manipulating with polynomials and high-level methods including polynomial GCD, factorization, and \Groebner bases over sophisticated ground rings. Special attention in the library is paid to high performance and a well-designed API. High performance is crucial for today's computational problems that arise in many research areas including high-energy physics, commutative algebra, cryptography, etc. \Rings'  performance is similar or even unmatched in some cases to many of advanced open-source and commercial software packages. The API provided by the library allows to write short and expressive code on top of the library, using both object-oriented and functional programming paradigms in a completely type-safe manner. 

Some of the planned future work for \Rings includes improvement of Gr{\"o}bner bases algorithms (better implementation of ``change of ordering algorithm'' and some special improvements for polynomials over $\QQ$), optimization of univariate polynomials with more advanced methods for fast multiplication, specific optimized implementation of $\GF(2, k)$ fields which are frequently arise in cryptography, and better built-in support for polynomials over arbitrary-precision real numbers ($\mathbb{R}[\vec X]$) and over 64-bit machine floating-point numbers ($\mathbb{R}64[\vec X]$).

\Rings is an open-source library licensed under Apache 2.0. The source code and comprehensive online manual can be found at \href{http://ringsalgebra.io}{http://ringsalgebra.io}.

\section{Acknowledgements}
The author would like to thank Dmitry Bolotin and Andrei Kataev for stimulating discussions. The author also would like to thank Takahiro Ueda for useful suggestions regarding benchmarks and Thomas Hahn for the help with improving the quality of  manuscript. The work was supported by the Russian Foundation of Basic Research grant \#16-32-60017, Russian Science Foundation grant \#18-72-00070 and by the FAIR-Russia Research Center.

\section{References}

\bibliographystyle{model1a-num-names}

\bibliography{bibliography} 

\end{document}